\documentclass[aps,prd,twocolumn,groupedaddress,showpacs,amssymb]{revtex4}
\usepackage[T1]{fontenc}
\usepackage[latin1]{inputenc}
\usepackage{graphicx}
\usepackage[english]{babel}
\usepackage{graphicx}
\usepackage{bm}
\usepackage{amsmath}
\usepackage{amssymb}
\usepackage{amsfonts}
\usepackage{epsfig}
\usepackage{colordvi}
\usepackage{color}
\def \rg {r_g}
\def \xns {x_{\rm ns}}
\def \xmax {x_{\rm max}}
\def \xcrit {x_{\rm crit}}
\def \rhoc {\rho_{\rm c}}
\def \Pc {P_{\rm c}}

\def \rhocrit {\rho^{\rm crit}_{\rm c}}

\def \mr {$M-\cal{R}$}

\begin{document}
\title{The Mass-Radius relation for Neutron Stars in $f(R)$ gravity}

\author{Salvatore Capozziello$^{1,2,3}$\footnote{e-mail address: capozzie@na.infn.it}, 
Mariafelicia De Laurentis$^{4,5,6,2}$\footnote{e-mail address: mfdelaurentis@tspu.edu.ru},
Ruben Farinelli$^{7}$\footnote{e-mail address: ruben.farinelli@gmail.com}, Sergei D. Odintsov$^{5,8,9}$\footnote{e-mail address: odintsov@ieec.uab.es}}
\affiliation{$^{1}$Dipartimento di Fisica, Universit\' a
di Napoli {}``Federico II'', Compl. Univ. di
Monte S. Angelo, Edificio G, Via Cinthia, I-80126, Napoli, Italy}
\affiliation{$^{2}$INFN Sezione di Napoli, Compl. Univ. di
Monte S. Angelo, Edificio G, Via Cinthia, I-80126, Napoli, Italy}
\affiliation{$^{3}$Gran Sasso Science Institute (INFN), Via F. Crispi 7, I-67100, L' Aquila, Italy}
\affiliation{$^{4}$Institut f\"ur Theoretische Physik, Goethe-Universit\"at,
  Max-von-Laue-Str. 1, 60438 Frankfurt, Germany}
\affiliation{$^{5}$Tomsk State Pedagogical University, ul. Kievskaya, 60, 634061 Tomsk, Russia}
\affiliation{$^{6}$Lab.Theor.Cosmology,Tomsk State University of Control Systems and Radioelectronics(TUSUR), 634050 Tomsk, Russia}
\affiliation{$^{7}$ISDC Data Center for Astrophysics, Universit\'e de Gen\`eve, chemin d'\'Ecogia 16, 1290 Versoix, Switzerland}
\affiliation{$^{8}$Instituci\`{o} Catalana de Recerca i Estudis Avan\c{c}ats (ICREA), Barcelona, Spain,}
\affiliation{$^{9}$Institut de Ciencies de l'Espai (IEEC-CSIC), Campus UAB, Torre C5-Par-2a pl, E-08193 Bellaterra, Barcelona, Spain}
\date{\today}

\begin{abstract}
We discuss the Mass -- Radius  diagram for static neutron star models obtained by the numerical  solution
of modified Tolman-Oppenheimer-Volkoff equations  in $f(R)$ gravity where the Lagrangians    $f(R)=R+\alpha R^2 (1+\gamma R)$ and $f(R)=R^{1+\epsilon }$ are adopted.
Unlike the case of the perturbative approach previously reported, the solutions
are constrained by the presence of an extra degree of freedom,  
coming from the trace of the field equations.
In particular, the stiffness of the equation of state  determines an upper limit
on the central density $\rhoc$ above which the the positivity condition 
of energy-matter tensor trace $T^{\rm m}=\rho - 3 p$
holds. 
In the case of quadratic f(R)-gravity, we find higher masses and radii at lower
 central densities with an  inversion of the behavior around a pivoting $\rho_c$ which
depends on the choice of the equation of state.
When considering the cubic corrections, we find solutions converging to
the required asymptotic behavior of flat metric only for $\gamma < 0$. A similar analysis is performed for $f(R)=R^{1+\epsilon }$
considering $\epsilon$ as the leading parameter. We work  strictly in the Jordan frame in order to consider matter minimally coupled with respect to geometry. 
This fact allows us to avoid ambiguities that could emerge in adopting  the Einstein frame. 
\end{abstract}



\maketitle

\section{Introduction}

The structure of a neutron star (NS) is strictly correlated with the equation of state (EoS), {\it i.e.} the relation between pressure
and density in its interior \cite{Lattimer}. Given an EoS, a mass-radius $(M-\cal{R})$ relation  and a corresponding maximal 
 mass can be derived, in principle, for any NS. Furthermore  the knowledge of   these parameters provide significant information related to the mechanism responsible for NS formation and possible effects on  the evolutionary history of their progenitors. For an introduction to the theory of relativistic stars, see for example \cite{psaltis}.

Up to now,  the physical properties of matter in strong gravity regimes can be uniquely  studied considering theoretical models  since it is not possible to produce similar environments in laboratory. Due to this situation, there are more than $100$ candidates for  EoS,  but only some them  should be reliable once observational probes fix them. Consequently, the measurement of NS masses is thus important for our understanding on the matter behavior in extreme regimes. Beside these considerations, it is  well known that Chandrasekhar, considering degenerate matter,   fixed  a theoretical  upper limit   for the stability of a non-rotating NS at  $1.44 M_\odot$  
\cite{Chandrasekhar}.
From an observational point of view, the
determination of mass can  be achieved  with accuracy  only for NS in binary systems. In particular, the most accurate mass measurements have been derived for the binary radio pulsars where  values of masses are around $1.35 M_\odot$  \cite{Thorsett}. However,  for  the X-ray pulsar Vela X-1, it  has been measured a 
a mass of the order $1.86\pm0.16 M_\odot$ \cite{Barziv,Rawls} and, for 4U
1822-371, a mass of the order  $2M_{\odot}$ \cite{Munoz}.
More massive NS were discovered, such as the millisecond radio pulsar $J0751+1807$ with  a measured  mass of $2.1\pm 0.2 M_\odot$ \cite{Nice}. Furthermore, the pulsar PSR $J1614-2230$ \cite{Demorest} has set  rigid constraints on various  EoS at strong density regimes.
In summary, the mass of a NS cannot exceed the maximal mass limit in the range  $3.2\div 3.6 M_\odot$ according to  General Relativity (GR) \cite{Rhoades,Nauenberg}. 
It is important to stress that  these  limits on maximal mass exclude many EoS according to the
observational  data. Therefore, since different assumptions provided different results,
we can say that the actual mass limit for NS
is still a mystery. We need  a more accurate derivation consistent with the observations. In other words, this state of art allows us to  assume that the NS mass should be in the range  $1.4M_\odot$ to $6M_\odot$. This situation is very unsatisfactory and
so there is a severe need for reliable methods to obtain the NS mass limit,
otherwise waiting for more precise observations. 

In this paper we will face the problem of NS mass  limit adopting $f(R)$ gravity. This is a straightforward extension of GR where one relaxes the strict request that the gravity action is linear in the Ricci scalar $R$ as in the Hilbert-Einstein case. These models can be seen as  simple cases of a more general class of    Extended Theories of Gravity \cite{PhysRepnostro,OdintsovPR,6,Mauro,faraoni,10,libroSV,libroSF}.  The reason why we adopt such an approach is that higher order curvature corrections can emerge in the extreme gravity regimes inside a NS \cite{asta,fiziev,doneva,stergio}. The effective pressure related to the curvature could naturally cure, in principle, some shortcomings of NS theory that are often addressed by asking for exotic EoS \cite{asta}. 
The philosophy of the present paper is to construct reliable $M-\cal{R}$ relations solving directly the full modified Tolman-Oppenheimer-Volkoff (TOV) system equations \cite{Oppenheimer}. In other words, in our numerical integrations, we are not  imposing arbitrary perturbation methods but solve numerically the full TOV system. In particular, we take into account quadratic and cubic corrections to the Ricci scalar and, in general, power-law $f(R)$ models. The aim is to control precisely how the results deviate from those of GR in order to see where strong gravity regimes emerge and affect the $M-\cal{R}$ relation.

The paper is organized as follows:
In Section \ref{due}, we derive  the modified TOV equations for $f(R)$ gravity.   In particular, we consider  $f(R)$ models with quadratic and   cubic corrections and generic power law $f(R)$ models.  Solutions of stellar structure equations are given in Section \ref{tre}.
In Section \ref{quattro}, we discuss the results  focusing, in particular,  on the $M-\cal{R}$ relation. Conclusions  and discussion are reported in  Section \ref{cinque}.
Through the paper, we will indicate with $\cal R$ the radius of the object and with $R$ the Ricci curvature scalar.

\section{Tolman-Oppenheimer-Volkov equations for $f(R)$ gravity}
\label{due}
Let us start from the  $f(R)$  action given by

\begin{equation}\label{action}
{\cal A}=\frac{c^4}{16\pi G}\int d^4x \sqrt{-g}\left[f(R) + {\cal L}_{{\rm
matter}}\right]\,,
\end{equation}
where $g$ is the determinant of the metric $g_{\mu\nu}$ and ${\cal L}_{\rm
matter}$ is the standard perfect fluid matter Lagrangian. The
variation of (\ref{action}) with  respect to  $g_{\mu\nu}$ gives
the field equations \cite{PhysRepnostro,OdintsovPR,6,Mauro,faraoni,10,libroSV,libroSF}:

\begin{equation}
\frac{df(R)}{d R}R_{\mu\nu}-\frac{1}{2}f(R) g_{\mu\nu}-\left[\nabla_{\mu} \nabla_{\nu} - g_{\mu\nu} \Box\right]\frac{df(R)}{dR}=\frac{8\pi G}{c^{4}} T_{\mu \nu },
\label{field_eq}
\end{equation}
where $\displaystyle{T_{\mu\nu}= \frac{-2}{\sqrt{-g}}\frac{\delta\left(\sqrt{-g}{\cal L}_m\right)}{\delta g^{\mu\nu}}}$ is the energy momentum tensor of matter.
Here we  adopt the signature  $\left(+,-,-,-\right)$. 
The metric for systems with spherical symmetry has the usual form 

\begin{equation}
    ds^2= e^{2w}c^2 dt^2 -e^{2\lambda}dr^2 -r^2 (d\theta^2 +\sin^2\theta d\phi^2),
    \label{metric}
\end{equation}
where $w$ and $\lambda$ are functions depending only on the radial coordinate $r$.
Within the star, matter is described  as a perfect fluid,  whose  energy-momentum tensor is $T_{\mu\nu}=\mbox{diag}(e^{2w}\rho
c^{2}, e^{2\lambda}p, r^2 p, r^{2}p\sin^{2}\theta)$,  where $\rho$ is the
matter density and $p$ is the pressure \cite{weinberg}. 
The equations for the stellar configuration are obtained adding the condition of hydrostatic equilibrium which can be derived from the contracted Bianchi identities

\begin{equation}
\nabla^{\mu}T_{\mu\nu}=0\,,
\label{bianchi}
\end{equation}
that give  the  Euler conservation equation 
\begin{equation}\label{hydro}
    \frac{dp}{dr}=-(\rho
    +p)\frac{dw}{dr}\,.
\end{equation}
From the metric \eqref{metric} and the field equations (\ref{field_eq}), it is possible
to derive the equations for the functions $\lambda$ and $w$ in the form \cite{capquark}

\begin{eqnarray}
\label{dlambda_dr}
\frac{d\lambda}{dr}&=&\frac{8 e^{2 \lambda } G \pi  r \rho }{c^2 \left(2\frac{df}{dR}+r R' \frac{d^2f}{dR^2}\right)}+\frac{e^{2 \lambda } \left[\left(r^2 R-2\right)
\frac{df}{dR}-f r^2\right]}{2 r \left(2 \frac{df}{dR}+r R' \frac{d^2f}{dR^2}\right)}\nonumber\\\nonumber\\&&
+\frac{\frac{df}{dR}+r\left[ \frac{d^2f}{dR^2}\left(2R'+r R''\right)+r R'^{2}\frac{ d^3f}{dR^3}\right]}{r \left(2 \frac{df}{dR}+r R' \frac{d^2f}{dR^2}\right)}\,,
\end{eqnarray}
and
\begin{eqnarray}\label{WR}
 \frac{dw}{dr}&=&\frac{8 e^{2 \lambda } G P \pi  r}{c^4 \left(2 \frac{d f}{dR}+r R' \frac{d^2f}{dR^2}\right)}
\nonumber\\\nonumber\\&&
+\frac{e^{2 \lambda } \left[f r^2+ \left(2-r^2 R\right)\frac{ df}{dR}\right]-2
\left(\frac{df}{dR}+2 r R' \frac{d^2f}{dR^2}\right)}{2 r \left(2 \frac{df}{dR}+r R' \frac{d^2f}{dR^2}\right)},\nonumber\\
\end{eqnarray}
respectively. In both Eqs. (\ref{dlambda_dr}) and (\ref{WR}), the prime denotes a derivative
with respect to  $r$ for the Ricci scalar $R(r)$.

The above equations are the modified TOV equations that, for $f(R)=R$, reduce to the standard  TOV equations of GR \cite{rezzollazan,landaufluid}. 
It is important to stress that,  in $f(R)$ gravity, the Ricci scalar is a dynamical variable and then we need a further equation to solve the system of equations (\ref{hydro}),  (\ref{dlambda_dr}) and (\ref{WR}). 
For this aim one needs to consider  the trace of the field Eqs. (\ref{field_eq}) which takes the form
\begin{equation}\label{trace_eq}
3\square \frac{df(R)}{dR}+\frac{df(R)}{dR}R-2f(R)=\frac{8\pi G}{c^4}(\rho-3p)\,,
\end{equation}
where
\begin{equation}\label{box}
\square=\frac{1}{\sqrt{-g}}\frac{\partial}{\partial x^{\nu}}\left(\sqrt{-g} g^{\mu \nu} \frac{\partial}{\partial x^{\mu}}\right)
\end{equation}
is the  d'Alembert operator in curved spacetime.
It can be easily checked that for $f(R)=R$, Eq. (\ref{trace_eq}) reduces to the trace of   GR, i.e. 

\begin{equation} \label{R0} R=-\frac{8\pi G}{c^4}(\rho-3p)\,.
\end{equation}
In order to close the system of Eqs. (\ref{hydro}), (\ref{dlambda_dr}),  (\ref{WR}) and
(\ref{trace_eq}),  one needs  to provide an EoS, $P(\rho)$ relating the
pressure and the density inside the star.

In the next subsection we will take into account two physically relevant  $f(R)$  Lagrangians  with the aim to   obtain  the  $M-\cal{R}$ diagram for NS.  We use a fully self-consistent non-perturbative approach to  solve 
the stellar-structure equation in their \emph{exact} form.

\subsection{The case  $f(R)=R+\alpha R^2 (1+ \gamma R)$ }
\label{sect_cubic}
\noindent
We first consider here a quadratic form of $f(R)$ with cubic corrections, according to

\begin{equation}
f(R)=R+\alpha R^2 (1+ \gamma R),
\label{fr_form_cubic}
\end{equation}
where both $\alpha$ and $\gamma$ have dimensions cm$^{-2}$.
This class of  models
can be  related to
the presence of strong gravitational fields and emerge, for example, in cosmology,
 to achieve inflation. In particular higher-derivative curvature terms   naturally  lead to the cosmic accelerated expansion of inflation
 \cite{PLANCK}. At  fundamental level,  their  origin
is related to the effective actions coming from quantum
gravity \cite{buchbinder} or from quantum field theory formulated in
curved spacetime \cite{birrell}. They  lead to renormalizable
models  at the one-loop level. In the extreme gravity regime 
 of NS, also if very far from the full quantum
gravity regime, it is realistic to suppose the emergence of   curvature corrections to  improve the
pressure effects. 
It is important to stress that such a model cannot be confronted with  the Solar System tests of GR since the quadratic and cubic terms emerge in strong gravity regime as discussed above. They cannot be present at Solar System scales since the only relevant term  in the weak field regime is the linear Ricci scalar $R$. The interest of this model, in the present context, is that the interior of a NS is a natural laboratory where high curvature regimes can emerge and lead the $M-\cal{R}$ relation (see also the discussion in \cite{asta}). In some sense, the interior of a NS is similar to the conditions of the early universe.

For this model, Eqs (\ref{dlambda_dr}) and (\ref{WR}) take the form


\begin{eqnarray}\label{tovlambda_alpha}
\frac{d\lambda}{dr}=\frac{4 e^{2 \lambda } G \pi  r \rho }{c^2 \left[1+R \alpha  (2+3 R \gamma )+r \alpha  (1+3 R \gamma ) R'\right]} \nonumber\\\nonumber
+\frac{1}{4 r \left[1+R \alpha  (2+3 R\gamma )+r \alpha  (1+3 R \gamma ) R'\right]}\times \\\nonumber 
 \{e^{2 \lambda } [R \alpha  \left(R \left(r^2-6 \gamma +2 r^2 R \gamma \right)-4\right)-2]  \nonumber\\ 
+2 [1+3 R^2 \alpha  \gamma +2 r \alpha  \left(R' \left(2+3 r \gamma  R'\right)+r R''\right)]\\\nonumber
  +4 R \alpha  [1+3 r \gamma  \left(2 R'+r R''\right)] \},
 \end{eqnarray}

and
\begin{eqnarray}\label{tovw_alpha}
\frac{dw}{dr}=\frac{4 e^{2 \lambda} G P \pi r}{[c^4 (1 + R \alpha (2 + 3 R \gamma) + r \alpha (1 + 3 R \gamma) R']}\\\nonumber
+\frac{1}{4 r [1 + R \alpha (2 + 3 R \gamma) + r \alpha (1 + 3 R \gamma)R']} \times\\\nonumber
e^{2 \lambda} [2 + 
   R \alpha (4 - R (r^2 - 6 \gamma + 2 r^2 R \gamma))]  \\\nonumber
   -2 [1 + R \alpha (2 + 3 R \gamma) + 
      4 r \alpha (1 + 3 R \gamma) R'],
\end{eqnarray}

while the trace Eq. (\ref{trace_eq}) becomes

\begin{eqnarray}\label{trace_alpha}
&&\frac{d^2R}{dr^2}=\frac{1}{6 c^4 r \alpha (1 + 3 R \gamma)}\times \\\nonumber&&
e^{2 \lambda} r [c^4 R ( R^2 \alpha \gamma-1) + 8 G \pi (-3 P + c^2 \rho)]\\\nonumber&&
-6 c^4 \alpha R' [3 r \gamma R' + (1 + 3 R \gamma) (2 + r w' - r \lambda')]
\end{eqnarray}



\subsection{The case  $f(R)=R^{1+\epsilon}$}
\label{sect_power}
Another interesting class is 
\begin{eqnarray}\label{LOGe}
f(R)=R^{1+\epsilon}\,,
 \end{eqnarray}
that are  power-law models.   If we assume small deviation with respect to GR, that is 
  $|\epsilon| \ll 1$,   it is possible to write
 a first-order Taylor expansion as

 \begin{eqnarray}\label{LOG}
R^{1+\epsilon}&\simeq & R+\epsilon R {\rm log}R +O (\epsilon^2) ,
 \end{eqnarray}
 which is  interesting in order to define the
right physical dimensions of the coupling constant and to
control the magnitude of the corrections with respect to
the standard Einstein gravity. 
A Lagrangian form like that in  Eq. (\ref{LOG}) has been widely tested  giving interesting results starting from Solar System up to cosmological scales.  Applications have been found in the case of the cosmological background of gravitational waves \cite{maurofelix}, or in comparing the effects of small deviations  on the apsidal motion of a sample of eccentric eclipsing detached binary stars \cite{mnrasfelix}. Very strong bounds have been worked out from null and timelike geodesics in the cases of Solar System \cite{barrow}. Furthermore, black holes solutions using also Noether symmetry approach have been found for this kind of Lagrangian  \cite{arturo,axially}. The interest of these models is related to the fact that the value of the parameter $\epsilon$ can straightforwardly  relate a weak field curvature regime $(\epsilon \simeq 0)$ to a  regime where strong curvature effects start to become relevant $(\epsilon \neq 0)$. In fact, as shown in \cite{barrow}, the Solar System constraints give essentially $\epsilon\rightarrow 0$. This is not the case for NS where high curvature regimes are well far from the Solar System weak field limits.
In this perspective,   $\epsilon$ could be different from zero in NS  and then probe deviations with respect to GR. In this sense, NS could constitute a formidable test for alternative gravity.

The explicit form of Eq. (\ref{dlambda_dr}) and (\ref{WR}) for the action (\ref{LOG}) 
takes the form

\begin{eqnarray}\label{tovlambda_power}
&&\frac{d\lambda}{dr}=\frac{8 e^{2 \lambda} G \pi r R \rho}{c^2 [2 R (1 + \varepsilon + \varepsilon {\rm log}R) + r \varepsilon R']}\\\nonumber&&
+\frac{1}{2 r R [2 R (1 + \varepsilon + \varepsilon {\rm log} R) + 
       r \varepsilon R']}\times\\\nonumber&&
\{e^{2 \lambda} R^2 [r^2 R \varepsilon - 2 (1 + \varepsilon + \varepsilon {\rm log} R)]\\\nonumber&&
+2 [R^2 (1 + \varepsilon + \varepsilon {\rm log} R) - 
    r^2 \varepsilon R'^2 + 
    r R \varepsilon (2 R' + r R'')]\},
\end{eqnarray}
and

\begin{eqnarray}\label{tovw_power}
&&\frac{dw}{dr}=\frac{8 e^{2 \lambda} G P \pi r R}{c^4 [2 R (1 + \varepsilon + \varepsilon {\rm log} R) + r \varepsilon R']}\\\nonumber&&
-\frac{1}{2 r [2 R (1 + \varepsilon + \varepsilon {\rm log} R) + r \varepsilon R']}\times \\\nonumber&&
\{e^{2 \lambda}
   R [r^2 R \varepsilon - 2 (1 + \varepsilon + \varepsilon {\rm log} R)] \\\nonumber &&
 +2 [R (1 + \varepsilon + \varepsilon {\rm log} R) + 
    2 r \varepsilon R']\},
\end{eqnarray}

while the trace equation is given by
   
   \begin{eqnarray}\label{trace_power}
  && \frac{d^2R}{dr^2}=\frac{R'^2}{R}+R'\left(\lambda'-\frac{2}{r}-w' \right)  \\\nonumber&&
   -\frac{e^{2 \lambda} R [c^4 R ( 1-\varepsilon) + 
      8 G \pi (3 P - c^2 \rho) + c^4 R \varepsilon {\rm log} R]}{3 c^4 \varepsilon}.
   \end{eqnarray}
   In the framework of these models, let us now discuss the stellar structure for this models with the aim to achieve the $M-\cal{R}$ relation for some physically relevant EoS.

\section{The stellar structure equations}
\label{tre}

Let us consider  dimensionless variables for solving the system of Eqs. (\ref{hydro})-(\ref{trace_eq}). We set
the definitions $x= r/\rg$, $R=R/rg^2$, $p=P/P_0$, $\rho=\rho/\rho_0$, where 
$\rg=G M_{\odot}/c^2$, $P_0=M_{\odot}c^2/r_g^3$ and $\rho_0=M_{\odot}/r_g^3$, where $\rg=1.48 \times 10^5$ cm.
By the substitution $R'=q$, the trace equation for $R$ is lowered by an
order of derivation, and the full set of equations  is reduced to a system of  first-order ODEs, which can be expressed as

\begin{eqnarray}
\lambda'= F_1(\lambda, w, w', R, q, q',p,x),\\\nonumber
w'= F_2(\lambda, \lambda', w', R, q, q',p,x),\\\nonumber
R'=F_3 (q,x),\\\nonumber
q'= F_4(\lambda, \lambda', w, w', R, q, p,x),\\\nonumber
p'= F_5(w', p,x).\\\nonumber
\end{eqnarray}
In such a way, we can deal with the stellar structure equations under the standard of dynamical systems (see also \cite{fiziev}).

The requirement of asymptotic  flatness for  the metric implies that 
$\lambda \rightarrow 0$, $w \rightarrow 0$, $R \rightarrow 0$ for $r \rightarrow \infty$.
The boundary value problem (BVP) can be reduced to an initial value problem (IVP)
by imposing initial conditions at the star center $x=0$ such that the function
asymptotically have the needed behaviour. 
For the pressure,  the condition is simply $p(0)=p_c$, where $p_c$ is determined
by the chosen EoS once the central density $\rho_c$ is fixed.
For the remaining functions, the problems deserve some considerations: first of all, we notice that the natural initial conditions for $\lambda$ and the Ricci scalar are
$\lambda(0)=0$ and $R'(0)=0$.
On the other hand, it is worth pointing out that the potential function $w$ appears
in Eqs  (\ref{hydro})-(\ref{trace_eq}) only through its first and second derivatives $w'$ and $w''$.
This means that its initial value $w(0)$ can be defined up to an arbitrary
constant. Once the full solution is obtained, the value of $w$ at the star center
can be shifted \emph{a posteriori} such that $w \rightarrow 0$ asymptotically.

The most critical point for the numerical solutions concerns the value of the Ricci scalar at the center.
In the classical version of the {\it shooting method}, the right initial value of  $R_0$ is given by 
the root of the function $F(R_0)=y(R_0, x_{\rm max})-R^b(x_{\rm max})$,
where $y(R_0, x_{\rm max})$ is the value of $R$ at the boundary of the domain
obtained with the initial condition $R_0$, while $R^b(x_{\rm max})$ is the
imposed boundary condition (BC).
Once provided an initial guess for $R_0$, algorithms for root-searching, such
as the Newton method  for quadratic convergence, give
the initial values, if they  are not too far from the  solutions.

The implicit assumptions of the shooting method however is that the function 
$y(R_0, x_{\rm max})$, and in turn $F(R_0)$, is defined for any $R_0$;
moreover the classical  computation requires to determine numerically 
the derivative $dy/dR_0$ at each iteration.

In our case, this approach has not been possible for two reasons: first of all, for some initial
guess values of $R_0$,  smaller than the right one, the function $R$ at some distance $x > \xcrit$
becomes negative, leading to unphysical solutions for  the system of Eqs (\ref{hydro})-(\ref{trace_eq}).
Additionally, the numerical computation of the derivative $dy/dR_0$ at $\xmax$ becomes  time-consuming 
and strongly ill-conditioned,  depending critically  on the choice of the step-size for $R_0$.
The latter problem originates because  of the diverging behavior of the negative $R$-function
at values progressively higher than $\xcrit$.

We thus implemented a different method as follows: we first define for $R_0$ an initial guess interval 
$R^{\rm min}_0 \approx R_0/5$ and $R^{\rm max}_0 \approx 5R_0$, where $R_0= 8\pi (\rhoc - 3 p_{\rm c})$
is the GR value.
Then we use a bisection method over the given interval, which
allows to get progressively  closer to the right value of $R_0$ through subsequent interval
subdivisions.  The iteration  stops when a further interval subdivision does not produce appreciable change at $R(\xmax)$.
The needed accuracy in the determination of $R_0$ depends on the form of $f(R)$ and its
variable parameters.
In the case of $f(R)=R+\alpha R^2 (1+ \gamma R)$, we found that, for all considered values of $\alpha$,
the fine-tuning needs to be very tight, independently of the value of $\gamma$.

However, the bisection procedure over the initial interval of $R_0$-values allows  to achieve a
precision as high as desired, which is in fact limited only by the numerical accuracy of the integration routines.
In general, a few tens of iterations are required leading to the determination of $R_0$
with a precision of order $\Delta R_0/R_0 < 10^{-7}$.
For the Lagrangian $f(R)=R^{1+\varepsilon}$, on the other hand, as $\varepsilon$ gets progressively
closer to zero, the results become much less sensitive to $R_0$; namely,  $R_0= k R^{\rm GR}_0$,
with $k$ of the order  2, gives the same stellar masses and radii.
In this case we considered the value $R_0=1/2(R^{\rm min}_0+R^{\rm max}_0)$.
The adimensional NS radius is defined by the  condition $p(x_{\rm ns})=0$;
for $x > \xns$ the integration is continued in vacuum for the metric potentials
$\lambda$ and $w$, and for the Ricci scalar $R$.

To check the accuracy of the results, we tested both a Gear $M=2$ method and $4^{\rm th}$-order Runge-Kutta  method with adaptive step-size and accuracy control of the functions during the  integration
of the set of above differential equations  (see \cite{algoritmi} for details); we found no significative differences
between the two methods. 
Finally, the gravitational mass is obtained a-posteriori by using a Gauss-Legendre quadrature
rule for the integral
\begin{equation}
M_{\rm grav}=\rho_0 \rg^3 \int^{\xns}_0 4 \pi x^2 \rho(x) dx,
\label{mgrav}
\end{equation}
where $\xns$ is the adimensional NS radius, while $\rho(x)$ is obtained by computing the inverse of the function  $p(\rho)$ of the EoS.

\section{Results}
\label{quattro}


As previously mentioned,  the stellar structure equations require to specify
the EoS relating the matter pressure and density. From a mathematical point of view, the EoS is the algebraic relation needed to close the system.
There is however an important point to keep in mind that
 has been surprisingly little discussed in literature, namely 
the maximal density $\rhoc$ at the center of the star which can be used in numerical simulations.

Depending on the stiffness of the EoS, the maximum implementable value of $\rhoc$ is given
by the need to preserve the physical condition $Tr(T)=\rhoc c^2 -3 \Pc > 0$ for the trace equation.
In the standard  TOV system  or in the perturbative approach, $Tr(T)$ does not appear explicitly
in the equations of the stellar structure. 
However in the exact treatment, $Tr(T)$ represents the source term in the right hand
side of the trace equation differential operator (see Eq. (\ref{trace_eq})). Values of $\rhoc$ violating its positivity lead to un-physical results. 
As a consequence, the top-left extension of the branches of \mr\ diagrams is limited
by the maximum achievable value of $\rhoc$ preserving the positivity of the energy-matter 
tensor for a given EoS. This limits of course the possibility to investigate 
the stability of compact stellar structures up to arbitrary central density values.
In this paper,  we used the analytical representation of EoS with different stiffness as reported in \cite{potekhin13} (labeled as BSk19,BSk20 and BSk21) and the Sly EoS 
reported in \cite{douchin2001}.
The maximum allowed central densities  in units of $10^{15}$ gr/cm$^3$ are
 $\rhoc=2, 1.25, 1.35, 1.7 $
for EoS BSk19,BSk20, BSk21 and Sly, respectively.

\subsection{Results for  $f(R)=R+\alpha R^2$}

\begin{figure}[h]
\begin{center}
  \includegraphics[angle=-90,width=0.5\textwidth]{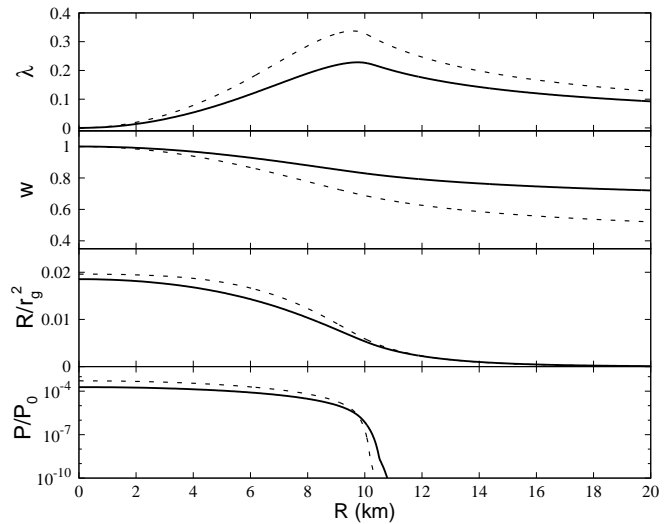}
  \caption{Radial profile for the metric potentials $\lambda$ and $|w|$, the dimensionless Ricci scalar $R/r^2_{\rm g}$ and the dimensionless pressure $P/P_0$, obtained from the solution of the system of  equations  (\ref{hydro}), (\ref{tovlambda_alpha}), (\ref{tovw_alpha}), and (\ref{trace_alpha}) for $f(R)=R+\alpha R^2$ with $|\alpha|=1$.  
  The central densities reported are $\rhoc=1 \times 10^{15}$ gr/cm$^3$
  (\emph{thick line}) and $\rhoc=1.5 \times 10^{15}$ gr/cm$^3$ (\emph{dashed line}), for the Eos BSk19.}
  \label{params_vs_r_cubic}
\end{center}
\end{figure}

We first considered quadratic $f(R)$ gravity as reported in Eq. (\ref{fr_form_cubic}) with
$\gamma=0$ and progressively increasing values
of $\alpha$. 
Note it has to be  $\alpha  < 0$ because of the adopted sign convection of the metric signature (see Eq. (\ref{metric})).
Adopting negative $\alpha$, we are avoiding ghost modes and instability behaviors \cite{ottewill}.
An example of the radial behavior of the metric potentials $\lambda$ and $w$, the Ricci scalar
$R$ and the pressure $P$  is shown in  Fig. \ref{params_vs_r_cubic}.

\begin{figure*}
\center
\includegraphics[angle=-90,scale=0.30]{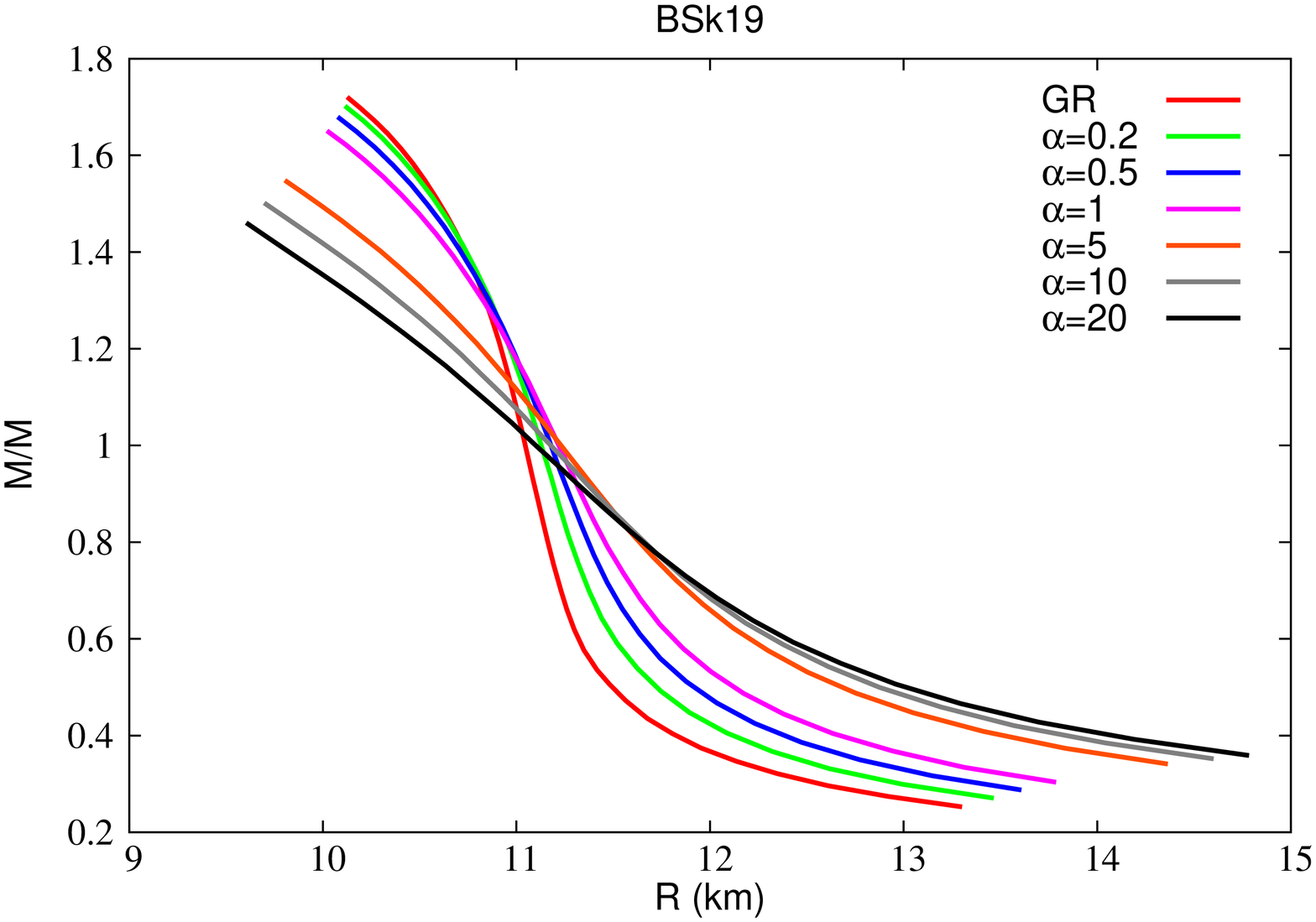}
\includegraphics[angle=-90,scale=0.30]{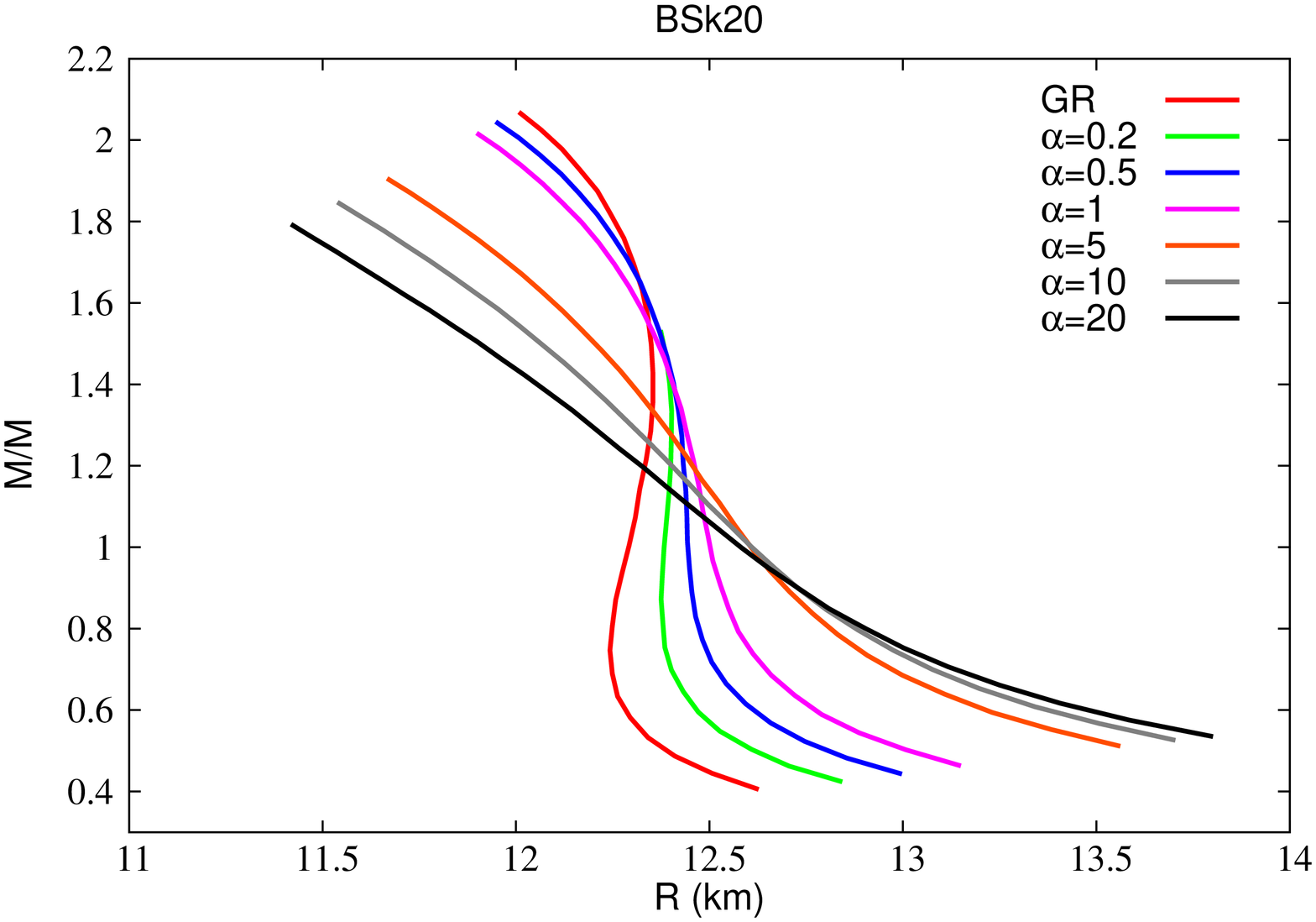}
\includegraphics[angle=-90,scale=0.30]{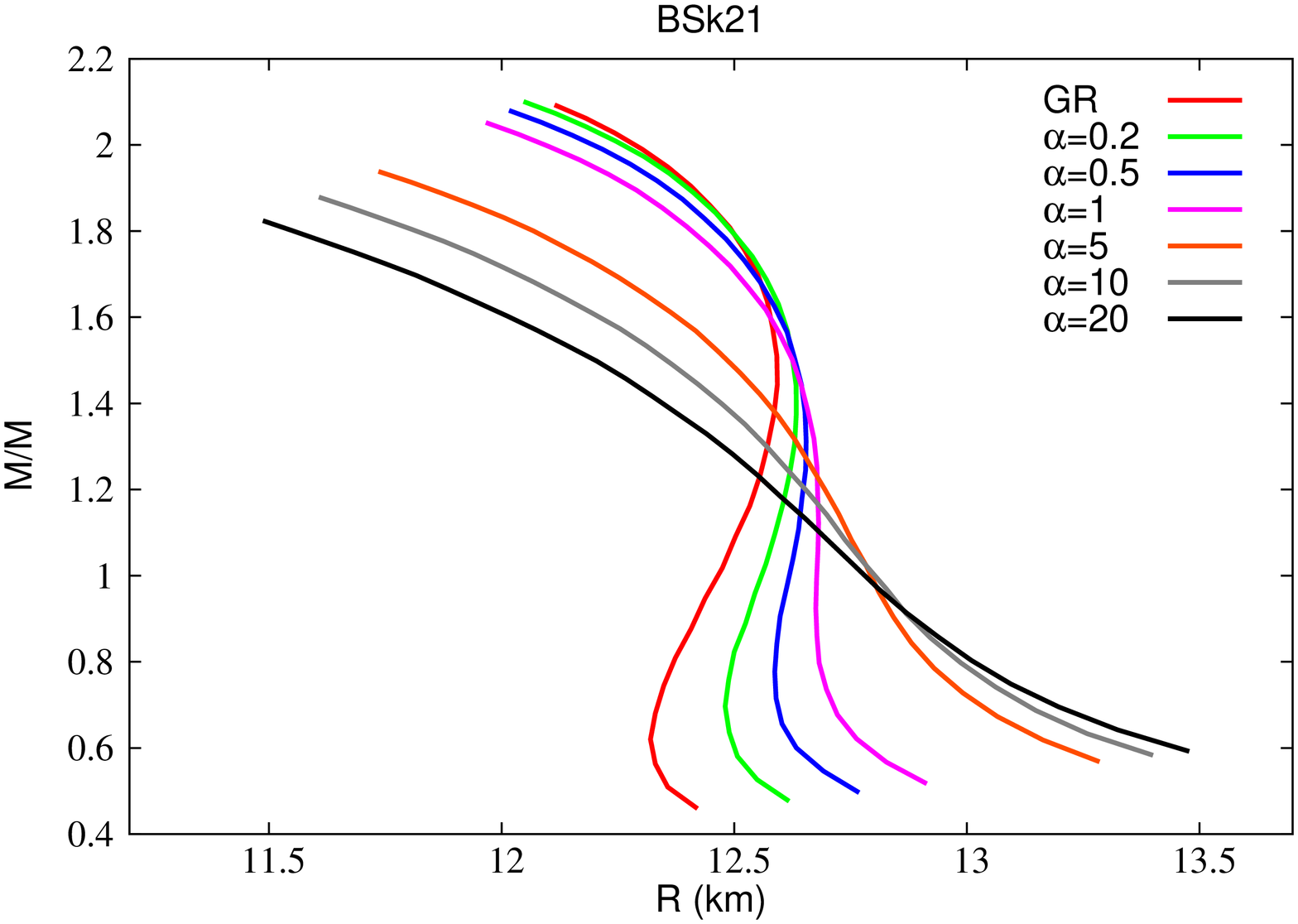}
\includegraphics[angle=-90,scale=0.30]{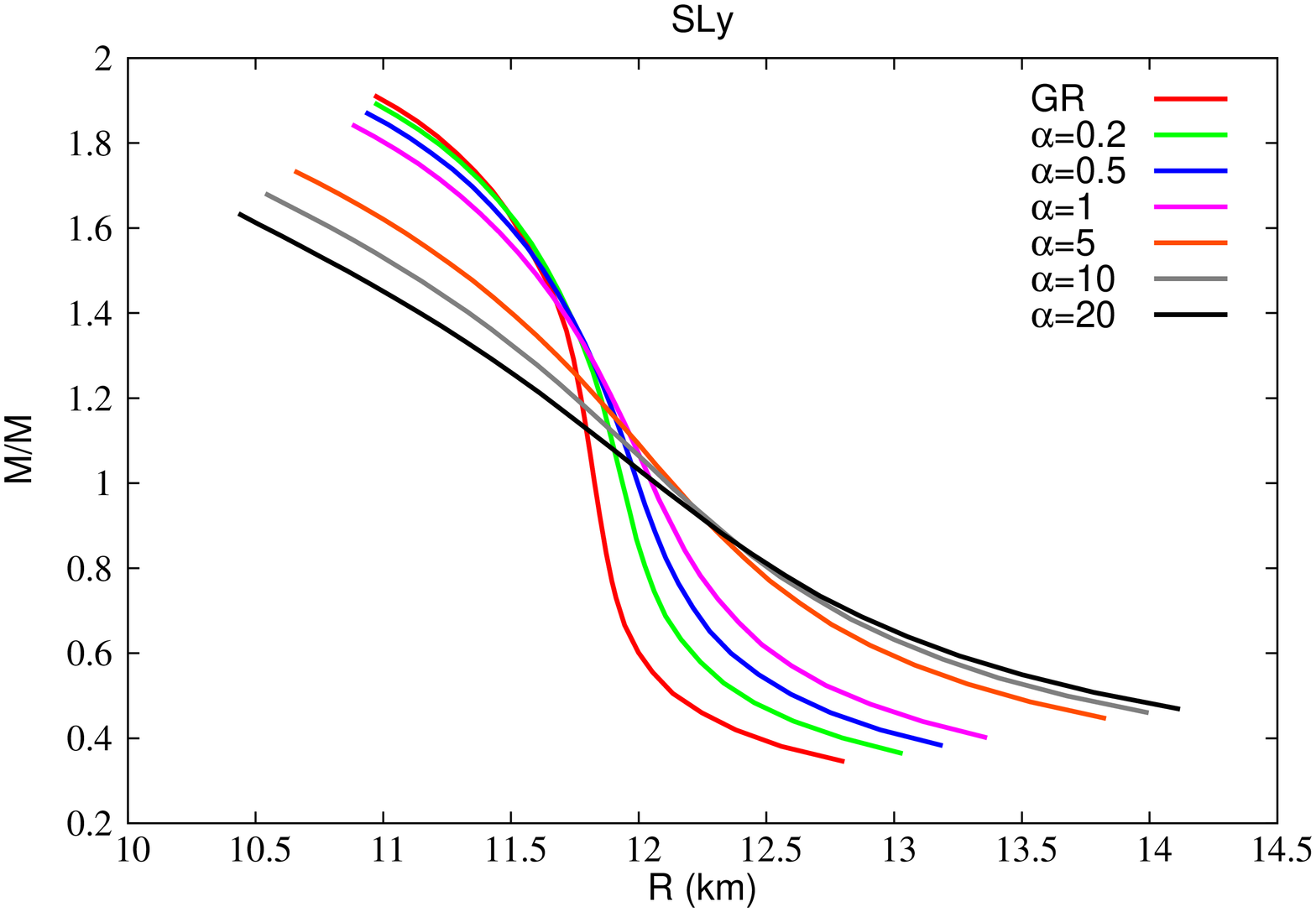} 
  \caption{\mr\ diagrams for different EoS (labeled in the top of each panel) and quadratic form  $f(R)=R+\alpha R^2$. For each EoS the maximal central density is determined by the condition $\rhoc -3 p > 0$ (see text). The classical TOV solution is also reported. The values of $\alpha$ are in modulus.}
  \label{fig_mr_quadratic}
\end{figure*}

\begin{figure*}
\includegraphics[angle=-90, scale=0.31]{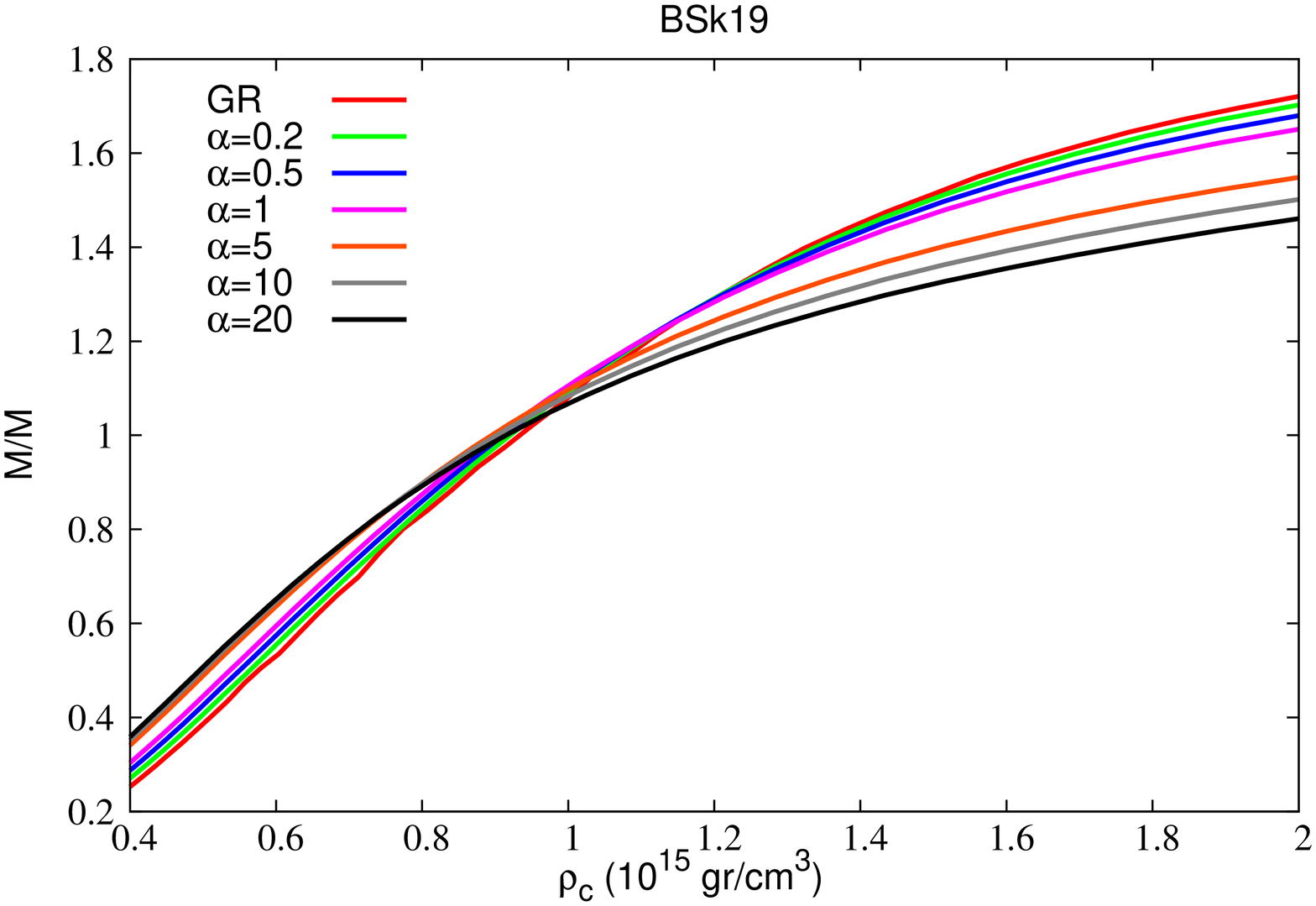}
\includegraphics[angle=-90,scale=0.31]{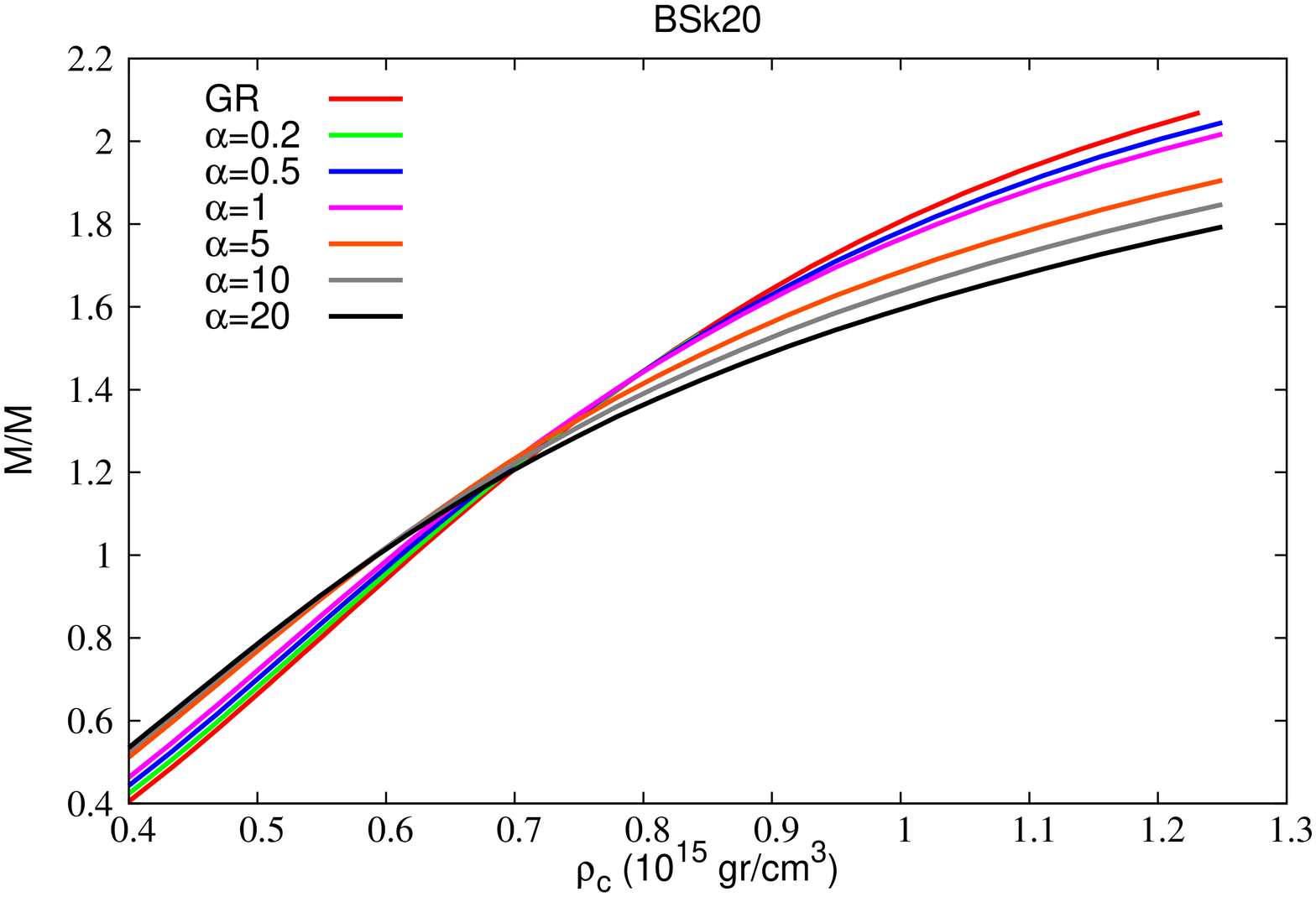} 
\includegraphics[angle=-90, scale=0.31]{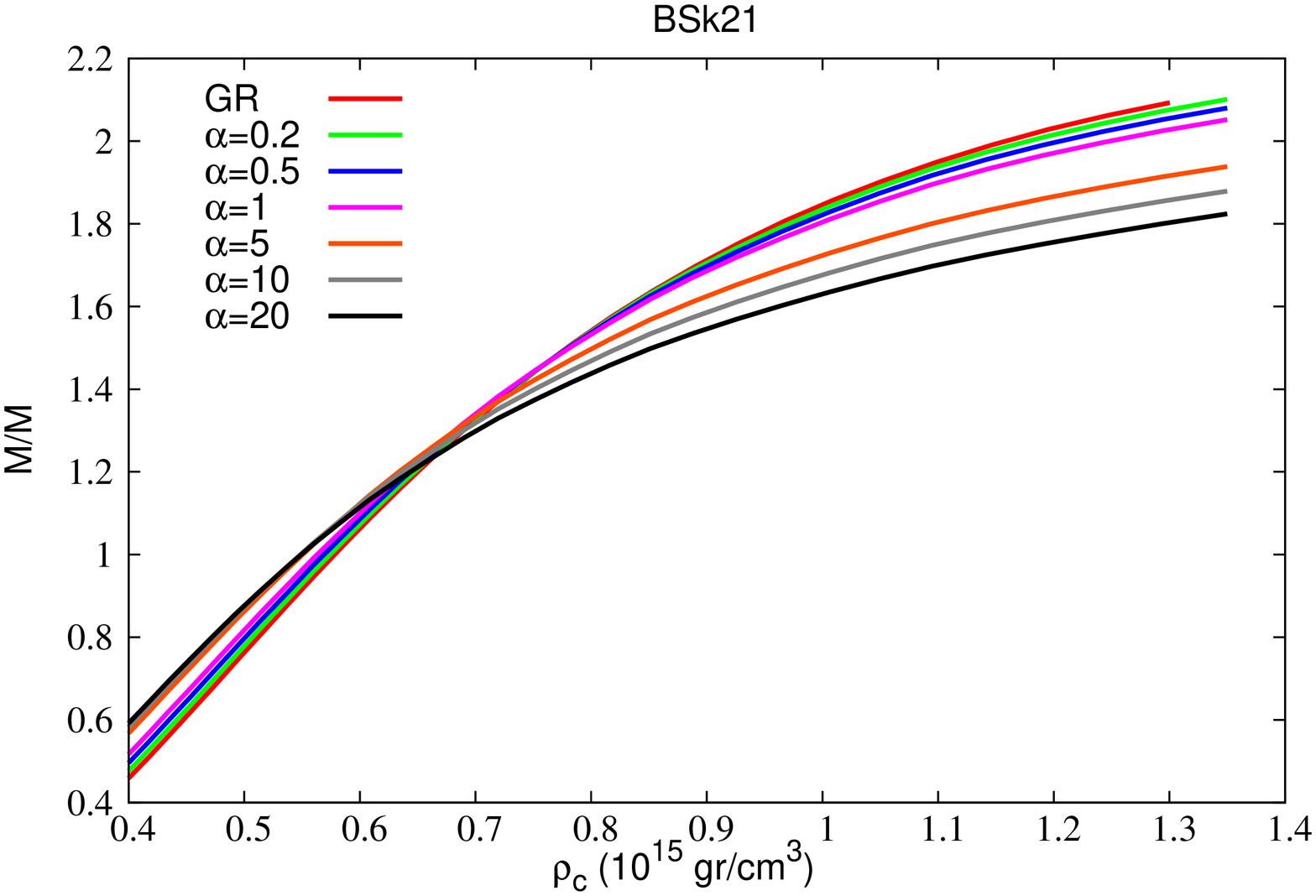}
\includegraphics[angle=-90,scale=0.31]{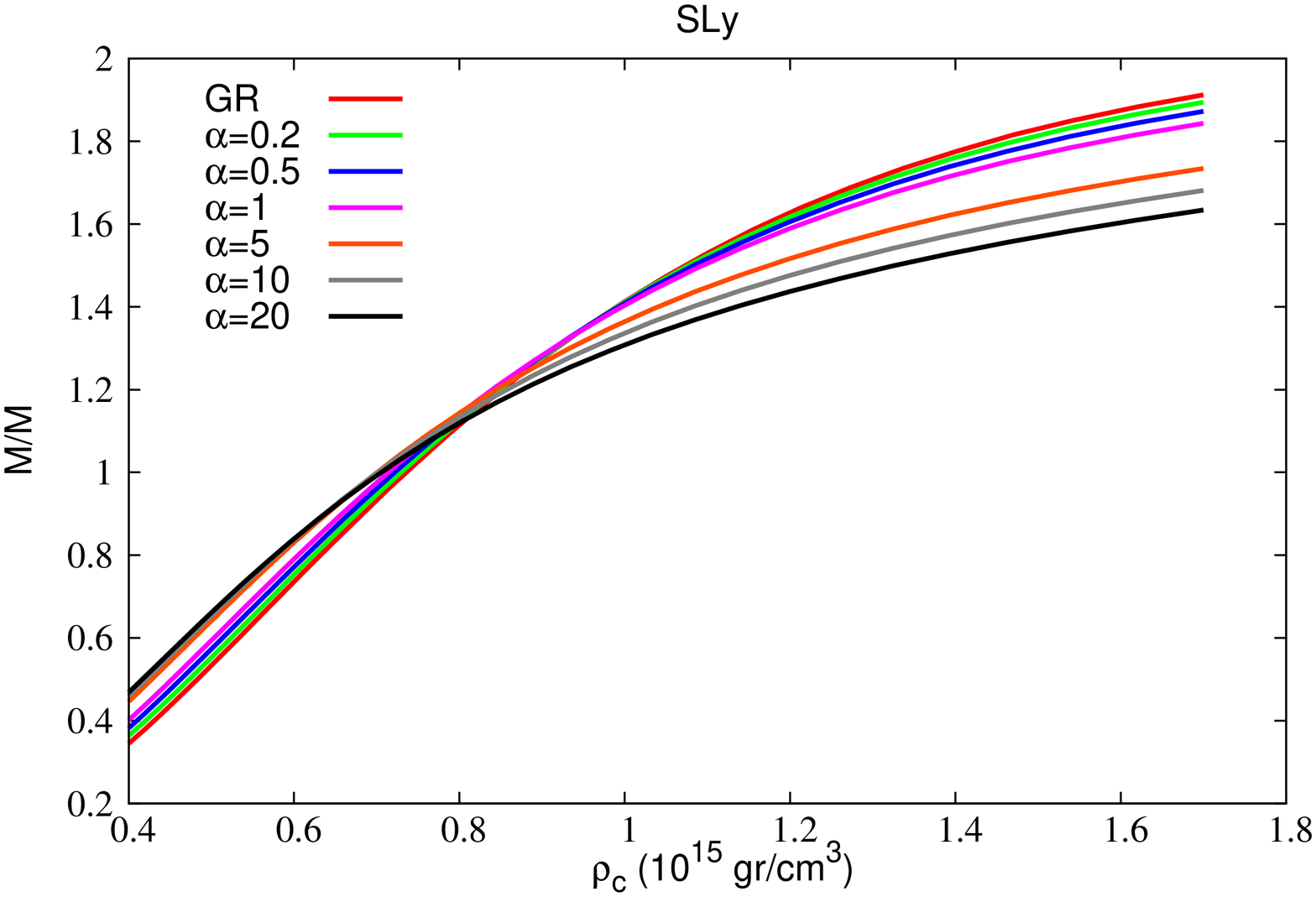} 
\caption{Mass versus central density for different EoS (labeled in the top of each panel) and quadratic form  f(R)=R+$\alpha$R$^2$. For each EoS the maximal central density is determined by the condition $\rhoc -3 p > 0$. As in Fig. \ref{fig_mr_quadratic}, the
modulus of $\alpha$ is reported (see text).}
  \label{fig_mrho_quadratic}
\end{figure*}

The \mr\  diagrams obtained for the four different EoS here considered are instead shown
in Fig. \ref{fig_mr_quadratic}, together with the solution of the classical TOV ($\alpha=0$). 
Independently of the EoS, as $|\alpha|$ increases,
the following behavior can be observed: the mass at lower central density are higher,
then the branch crosses the classical GR solution and for $\rhoc > \rhoc^{\rm crit}$ 
$M_{f(R)} < M_{\rm GR}$. The value of $\rhocrit$ depends on the EoS, and it is approximatively 
$0.6 \times 10^{15} < \rhocrit < 1 \times 10^{15}$ (Fig. \ref{fig_mrho_quadratic}).
The results reported in Fig. \ref{fig_mr_quadratic} and Fig. \ref{fig_mrho_quadratic} are obtained for values 
up to $|\alpha|=20$. Further higher values produce additional \emph{counter clockwise} rotation
of the $M-\cal{R}$ traces, until getting a saturation for $|\alpha| > 100$. 
There is an anti-correlation between $\alpha$ and the required value 
of the Ricci scalar $R_0$ at the NS centre simultaneously satisfying the asymptotic condition $R \rightarrow 0$.
This can be qualitatively understood also looking at the trace of the field equations in quadratic $f(R)$-gravity, which can be written as
\begin{equation}
\Box R+ m^2 (R+\chi T)=0,
\label{trace_quad}
\end{equation}
where ${\displaystyle m^2=-\frac{1}{6 \alpha}}$ is an effective mass, ${\displaystyle \chi=\frac{8\pi G}{c^4}}$ and, in vacuum,  the $R$-function exponentially decays approximatively
as $e^{- m r}$. The higher the value of $|\alpha|$, the higher is the distance at which
the Ricci scalar asymptotically goes to zero.
A function obeying the condition $R'(0)=0$ and the asymptotic zero-value condition at large radii needs to be smaller  at $r=0$ when the exponential-decay constant $m$ decreases.

\subsection{Results for  $f(R)=R+\alpha R^2 (1+ \gamma R)$}

\begin{figure}
\includegraphics[angle=-90,scale=0.30]{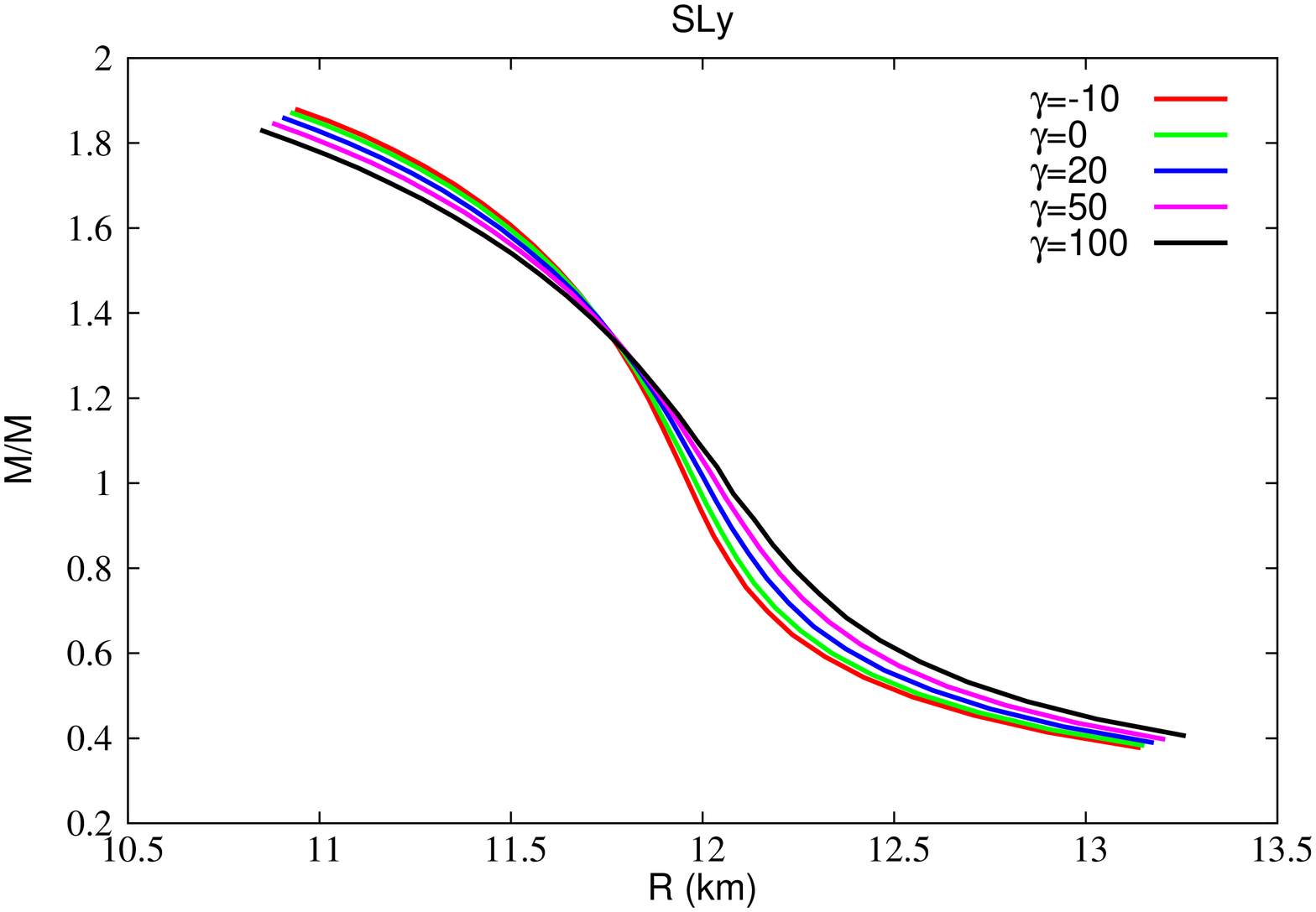} 
  \caption{\mr\ diagram for $f(R)=R+\alpha R^2 (1+\gamma R)$ with $|\alpha|=0.5$ and different values of $\gamma$, and Sly EoS.}
  \label{fig_mr_cubic}
\end{figure}

Let us improve now the above considerations by taking into account  $f(R)$ models  with non-zero  cubic correction term with
$\gamma\neq 0$ in the Lagrangian (\ref{fr_form_cubic}).
It is also evident that the higher the value of $|\alpha|$, the higher
must be the value of  $\rvert\gamma \rvert$ in order to see appreciable 
deviations from  quadratic $f(R)$-gravity.
We discuss here the case $|\alpha|=0.5$ as an example, showing the results
in Fig. \ref{fig_mr_cubic} for the case of a SLy EoS. 
For positive and progressively increasing values of $\gamma$, the behavior of
the traces in the $M-\cal{R}$ diagram is similar to that of the quadratic $f(R)$-form
at increasing values of $|\alpha|$ -- higher masses and radii below a certain
central density $\rhoc$ and inversion of the trend above it.
This results can be understood by keeping in mind the definition of $f(R)$ in
Eq. (\ref{fr_form_cubic}), where the cubic term  $\alpha\gamma R^3$ is negative
for $\gamma > 0$ (being  $\alpha <0$ because of the adopted signature). 

Thus, positive $\gamma$-values provide a further decrease of the value of the
Lagrangian density $f(R)$ in the same fashion as the quadratic coefficient $|\alpha|$ does.
This negative contribution unavoidably reflects in the $M-\cal{R}$ relation both for  the quadratic and quadratic plus cubic  corrections in  $R$.

At the opposite, negative $\gamma$-values  produce a \emph{clockwise}
rotation of the $M-\cal{R}$ trace. This can be seen indeed in Fig. \ref{fig_mr_cubic}
for the case $\gamma=-10$.
We found however that if from one side $\gamma$ can take arbitrary high
negative values, this is not true in the opposite case.
For instance, for the case $|\alpha|=0.5$ and $\gamma=20$, here discussed,
we were not able to find converging solutions matching the required asymptotic
behavior. 
As outlined  above, the critical maximum value of positive $\gamma$-values
depends however on the simultaneous choice of $\alpha$. 
As an example, the combination $|\alpha|=5 \gamma=20$ gives rise to a converging
solution which is however almost indistinguishable from the case with $\gamma=0$.

In general, we can say that for each value of $|\alpha|$, there is a 
critical positive value $\gamma^{\rm crit}$ above which solutions
are not allowed. More specifically, for $0 \gtrsim \gamma \gtrsim \gamma^{\rm crit}$
quadratic and cubic forms of $f(R)$ provide very similar results, while
for $\gamma < 0$ it is possible to go beyond values where modifications
of the $M-\cal{R}$ relation are more evident.
In the perturbative approach  both positive and negative values of $\gamma$ were considered  \cite{asta}.

\subsection{Results for  $f(R)=R+\epsilon R {\rm log} R$ }

\begin{figure}
\includegraphics[angle=-90,width=0.5\textwidth]{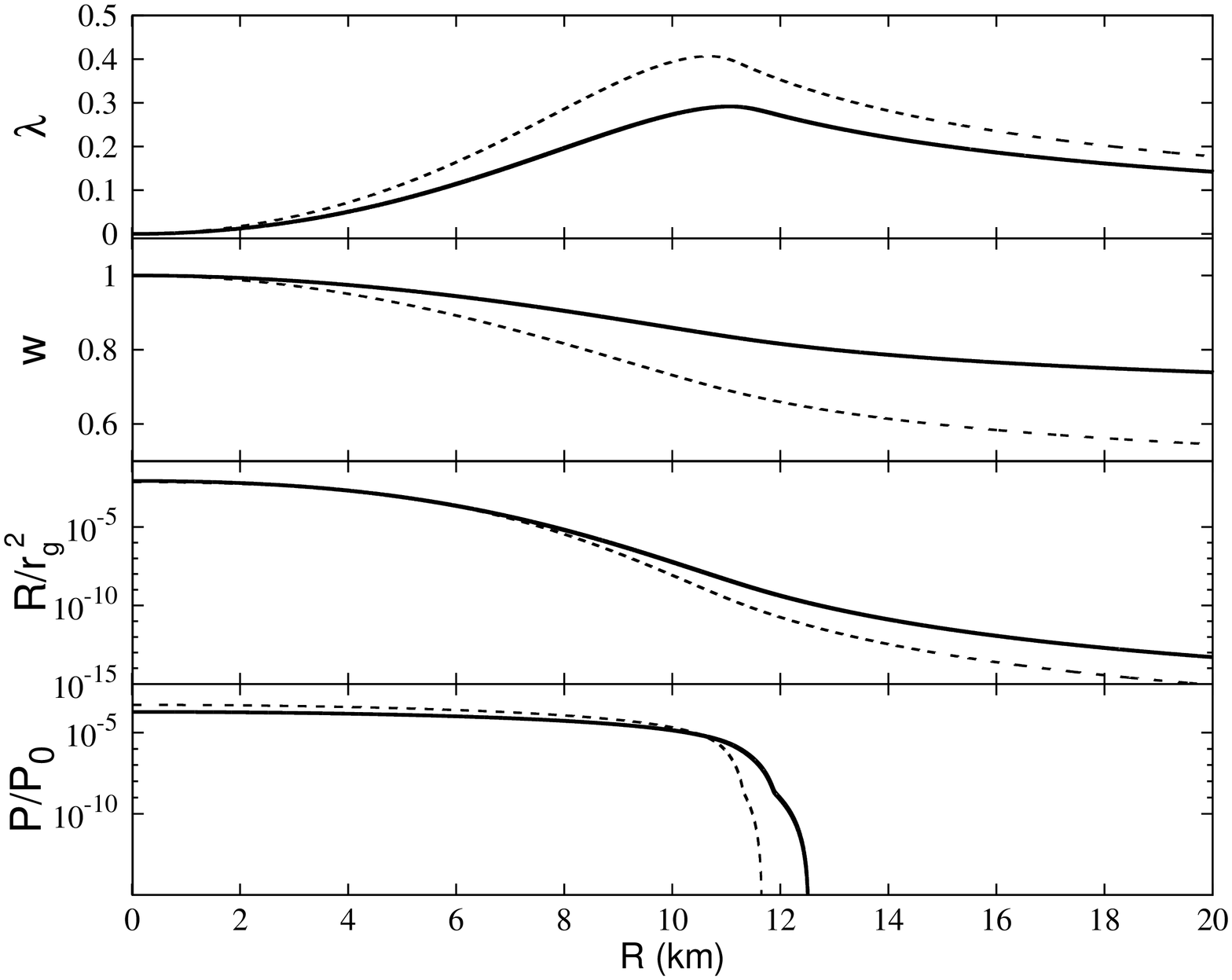} 
  \caption{Same as in Fig. \ref{params_vs_r_cubic} but for $f(R)=R+\varepsilon R {\rm log}R$ with $\varepsilon=-0.05$.}
  \label{params_vs_r_power}
\end{figure}

\begin{figure*}
\includegraphics[angle=-90,scale=0.30]{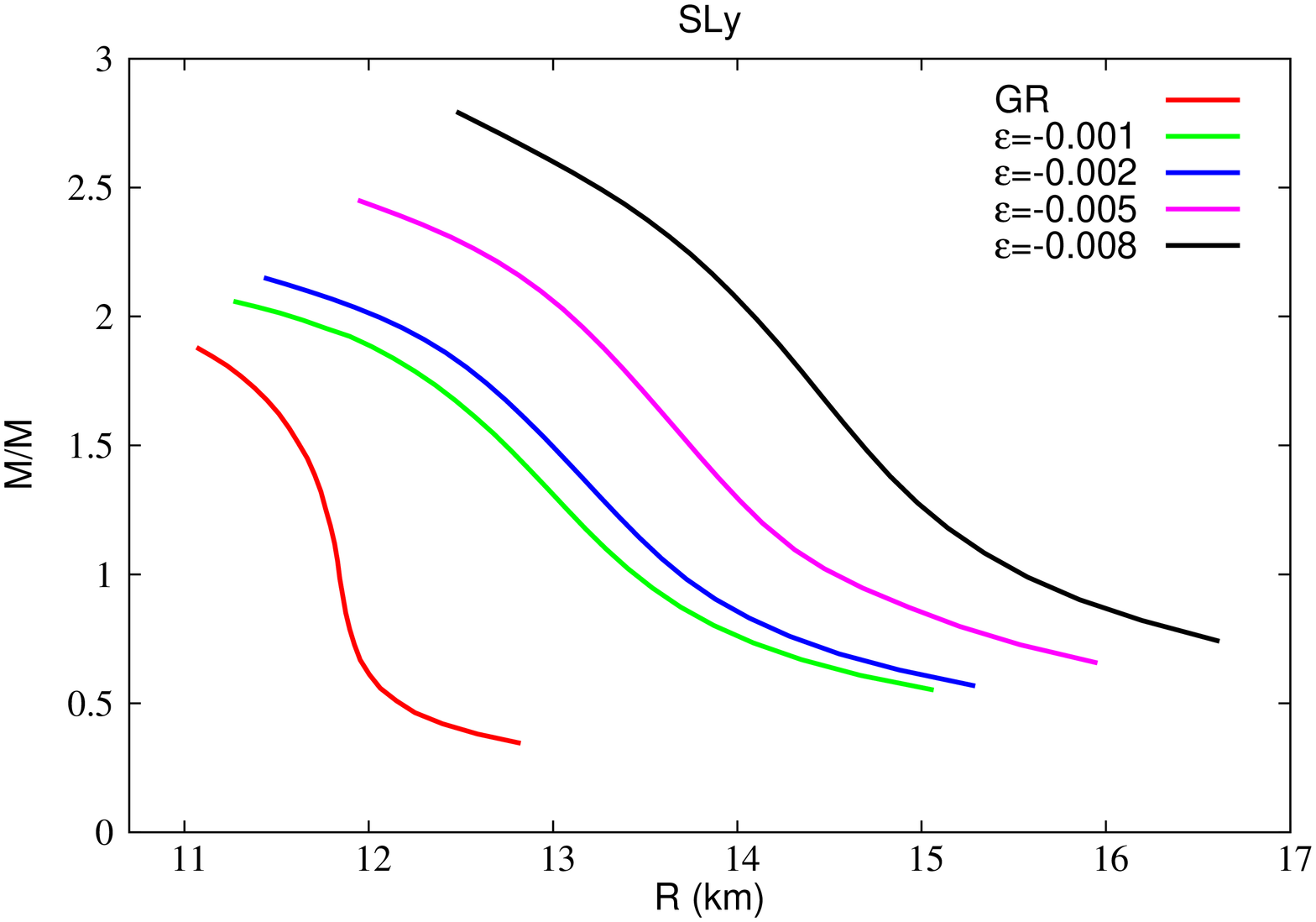}
\includegraphics[angle=-90,scale=0.30]{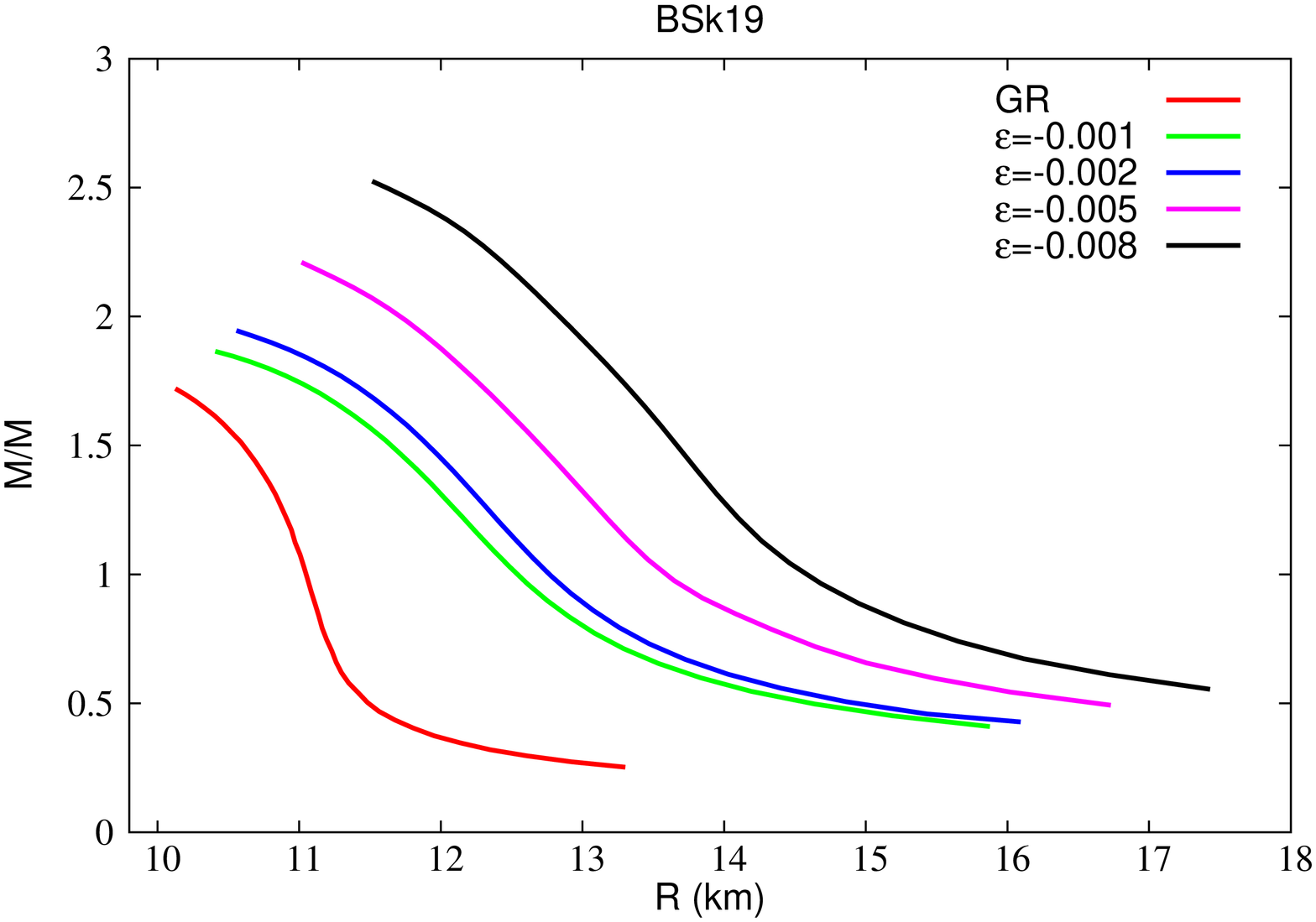}\\
\includegraphics[angle=-90,scale=0.30]{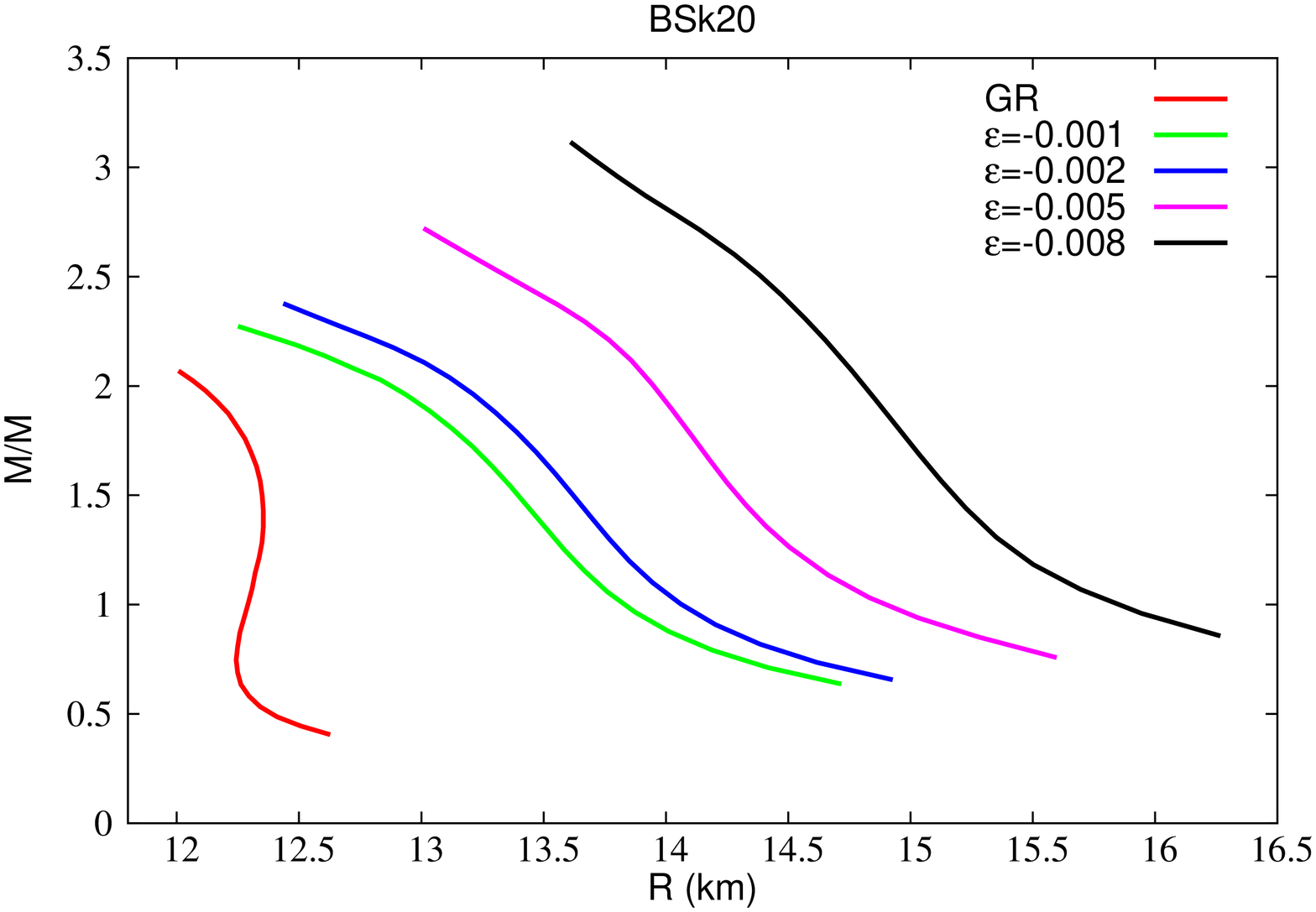}
\includegraphics[angle=-90,scale=0.30]{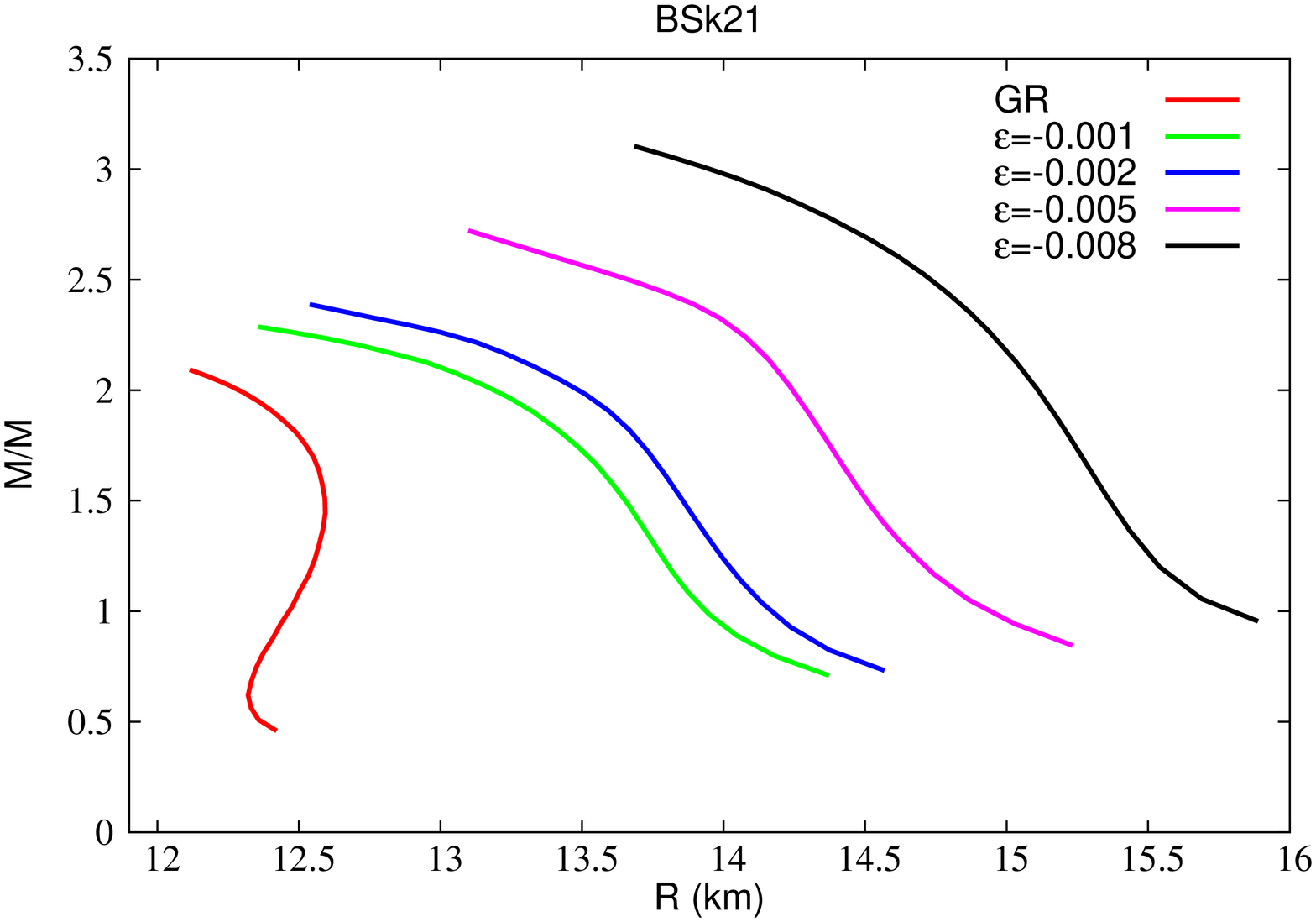}\\
\caption{\mr\ diagrams for  $f(R)$ given in equation (\ref{LOGe}) with four EoS and different values of $\epsilon$. The classical TOV corresponding to $\epsilon =0$ is also reported.}
 \label{fig_mrpower}
\end{figure*}

Finally,  we report the  results for the model  given in Eq. (\ref{LOGe}).
Albeit the field equations given in Sec. \ref{sect_power} are presented for clarity using the
Taylor expansion for $f(R)$ of Eq. (\ref{LOG}), it is worth noticing  that we actually
solved the stellar structure  equations  using the exact form $f(R)=R^{1+\varepsilon}$.
Similarly to Fig. \ref{params_vs_r_cubic}, we show, in Fig. \ref{params_vs_r_power},   an example of the radial behavior of the functions $\lambda$, $w$, $R$ and $P$.
The reason why we  considered the case $\varepsilon \ll 1$ can be better
understood looking at the \mr\ diagram presented in Fig. \ref{fig_mrpower}. 
For $|\varepsilon| < 0.01$ a significant deviation is observed with respect to the classical
TOV; at first sight, it is noticeably that the traces in the diagram present a self-similar
behavior for increasing values of $|\varepsilon|$. The results of physical interest (higher masses  and radii) are obtained for $\varepsilon < 0$, while, in the case of positive values of $\varepsilon$, the traces are blended with respect to the classical TOV solution. Note that in this case there is no pivoting around some critical central
 density, unlike what found in the case of the polynomial form of $f(R)$.
 Actually the resizing of the action $f(R)=R^{1+\varepsilon}$ reflects in a resizing of the
 traces in the \mr\ diagrams.
 In the case of Eos BSK20 and BSK21, NS masses,  larger than$ 3 M_{\odot}$, can be easily achieved; these  values must also be considered as lower limits on the NS mass, which can be even higher for fast  rotating objects.
\section{Discussion and Conclusions}
\label{cinque}
Neutron Stars have a main role in relativistic astrophysics for several reasons. They are the most stable compact objects of the universe (a part the black holes) where matter reaches extremely high field regimes. Besides, they are very important in order to understand the final stages of stellar evolution. It is important to point out that the extreme field   regimes in NS cannot be achieved in any ground-based lab so they have acquired a crucial  role also in nuclear and particle physics.  Due to this feature, understanding the EoS working into a NS has a twofold meaning: from one side, it can give information on the state of matter in this compact objects, on the other side, it is relevant at astrophysical level to understand the global behavior of stars in the final stages of their life. 
Besides, these standard roles, NS could be extremely important to test alternative theories of gravity, due to the huge gravitational field acting on them.

The aim of this paper is to show that the \mr\ relation of NS can be consistently achieved by extended theories of gravity as $f(R)$ gravity.  In particular, we proposed some physically relevant $f(R)$ models and modified the TOV equations accordingly. In the  stellar structure equations, new terms related to curvature corrections come out and lead the evolution of the \mr\ relation. Physically, these terms assume the role of a sort of curvature pressure capable of leading the mass and the radius of the star \cite{asta}. Specifically, we reduced the stellar structure equations to a dynamical system and integrated it numerically putting in evidence the role of curvature corrections into  the integration. The resulting \mr\ diagrams strictly depends on the value of the curvature corrections, the sign of the correction parameters, and the chosen EoS. We dealt with these terms as corrections to the  GR in order to control deviations with respect to the standard Einstein theory. 

This point deserves a further discussion according to the numerical results presented in the above figures.  Let us consider first Fig.\ref{fig_mr_quadratic}, where the \mr\ relation is reported for several   values of the parameter $\alpha$  and different EoS. The GR value is for $\alpha=0$ while $\alpha\neq 0$ represents  corrections with respect to GR. Clearly the increasing of $\alpha$ in modulus gives rise to a sort of stretching-blending rotation  of the \mr\ relation slope around a fixed 
\mr\ point that sits in  the intervals $(1\div1.2) M_\odot$ and $(11\div12.5)$ km. This means that the magnitude of the gravitational corrections alters the structure of the NS thanks to the further curvature pressure-density terms present in the TOV equations (see also \cite{asta}). In some sense, given a EoS, curvature corrections result relevant in shaping the stellar structure and determining the \mr\  relation. Specifically,   the stretching-blending rotation depends on the effective pressure related to the curvature corrections in Eq. (\ref{dlambda_dr}) and, in particular in Eq. (\ref{tovlambda_alpha}). Increasing the parameter $\alpha$ means that the original \mr\ relation of GR is affected by a further pressure term that modifies the effective mass of the star and consequently is radius. However, also the effective density results modified by the curvature corrections. Then the net effect is the  stretching-blending rotation around $M$ and $\cal R$ values where curvature corrections are not so relevant (see Fig.\ref{fig_mr_quadratic}). It is important to stress again that the stretching-bending  effect is due to the curvature corrections and not to the change of  standard matter EoS.

Similar considerations hold also for the $M-\rho_c$ diagrams (Fig.\ref{fig_mrho_quadratic}) where the total mass of the NS is given as a function of the central density $\rho_c$.  Also in this case we have a stretching-blending rotation around given values of mass and density (GR values) depending on the absolute value of $\alpha$. Inserting  also the cubic correction,  led by the parameter $\gamma$, the trend is similar (see Fig.\ref{fig_mr_cubic}).

The case of $R^{1+\epsilon}$ gravity is totally different. In this case, as we can see from Fig.\ref{fig_mrpower}, different values of the parameter $\epsilon$ make to scale   the \mr\ relation so that larger stable structures, in terms of mass and radius, are achieved also if slight variation with respect to the value  $\epsilon=0$ are considered. This fact could be extremely relevant in order to address extreme massive NS  as revealed by some observations \cite{asta}. Specifically, while for the quadratic and cubic models, the \mr\ relation is only modified (stretching-blending rotation) just changing the stability region of  NS with respect to GR, here the curvature pressure  and  density give rise to a scaling law. This is quite obvious due to the $R^{1+\epsilon}$ model, but the physical consequences are  relevant since extreme massive and large objects can be achieved, as one can see from Fig. \ref{fig_mrpower}. In such a case, the breaking of GR behavior, related to $\epsilon\neq 0$, gives rise to a new stability branch for NS allowing extreme objects. We stress again that this phenomenon is strictly related only to the effective curvature quantities and not to the change of   standard matter EoS.

It is important to stress that we never used exotic matter but only realistic EoS.  This point is crucial since we adopted always the Jordan frame so that the gravity sector results corrected while the matter sector is unaffected. In such a way, the geodesic structure is not altered and standard EoS can be assumed. On the contrary, if we were adopting the Einstein,  frame,  we would have the standard gravity sector but a non-minimally matter sector. In such a case, geodesic structure results altered and it could be dangerous to adopt standard EoS. In other words, as it is shown in detail in  \cite{stabile}, the conformal transformations have the effect of shifting the non-minimal coupling (in our case $f'(R)^{-1}$) from the gravitational to the matter sector and then the meaning of quantities like gravitational potentials, pressure,  matter density and mass result more complicated and have to be accurately discussed.
In summary, we develop our calculations in the Jordan frame considering it as the "physical" frame and avoiding the ambiguities that could emerge in the Einstein frame
(see, for example, \cite{bricese, troisi}). The same approach is adopted also in \cite{ruben}. 

As an example of these considerations, we can see that,
qualitatively, the behavior of the traces in the \mr\ diagrams with respect to GR is the opposite with respect to  that reported by \cite{yazadjiev14} for the same form $f(R)=R+\alpha R^2$ (see Fig.\ref{fig_mr_quadratic} and Figs.1 and 2 in \cite{yazadjiev14}). The only possibility to explain this crucial differences relies  on the fact that computations in \cite{yazadjiev14} were performed in the Einstein frame, while in our paper we work in the Jordan frame.  The only way to compare exactly the results is to compare the behavior of the \mr\ diagram under conformal transformations assuming the same EoS. This will be the topic of a next project. Finally, the results of this work will be extended to magnetic as well as rotating neutron stars.

\appendix
\section{ Analytical representations of Equations of State}
Here we report the functional form of EoS that we  used along the paper to solve numerically the stellar structure equations.
The pressure can be parameterized as a function of density. Let us denote with $\xi=\log(\rho/\textrm{g\, cm}^{-3})$ the dimensionless density and with 
$\zeta = \log(P/dyn\, cm^{-2})$ the dimensionless pressure. We  used the SLy equation as reported in \cite{hp2004} for non-rotating NS configurations. 
  It is 
\begin{eqnarray}
  \zeta &=& \frac{a_1+a_2\xi+a_3\xi^3}{1+a_4\,\xi}\,f_0(a_5(\xi-a_6))
\nonumber\\&&
     + (a_7+a_8\xi)\,f_0(a_9(a_{10}-\xi))
\nonumber\\&&
     + (a_{11}+a_{12}\xi)\,f_0(a_{13}(a_{14}-\xi))
\nonumber\\&&
     + (a_{15}+a_{16}\xi)\,f_0(a_{17}(a_{18}-\xi)) ~\,,
\label{eq:fit.P}
\end{eqnarray}
where the parameters $a_i$ for SLy EoSs are given in
Table~\ref{tab:fit.P1}.
\begin{table}[ht!]
\centering
\caption[]{Values of $a_i$(SLy) parameters for Eq. (\protect\ref{eq:fit.P}).}
\label{tab:fit.P1}
\begin{tabular}{rl||rll|}
\hline\hline
\rule[-1.4ex]{0pt}{4.3ex}
i  & $a_i$(SLy) & i & $a_i$(SLy)  \\
\hline\rule{0pt}{2.7ex}
$1$     &~\,${6.22}$  &  ${10}$ &~\, 11.4950\\    
$2$     & ~\,${6.121}$  & ${11}$ &~\, $-22.775$ \\
    $3$     &~\, $0.005925$ & ${12}$ &~\, 1.5707 \\     
$4$     & ~\,$0.16326$ &  ${13}$  &~\, 4.3    \\ 
$5$     &~\, $6.48$   &   ${14}$  &~\, 14.08  \\
$6$     &~\, 11.4971 &  ${15}$ &~\, 27.80  \\
$7$     &~\, 19.105   &  ${16}$ &~\, $-1.653$  \\
$8$     &~\, 0.8938  &  ${17}$ &~\, 1.50   \\
$9$     &~\, 6.54   &   ${18}$ &~\, 14.67  \\
 \rule[-1.4ex]{0pt}{0pt}\\
\hline\hline
\end{tabular}
\end{table}
\begin{table}[ht!]
\centering
\caption[]{Numerical values of $a_i$ parameters for the Eq.\eqref{fit.P}.}
\label{tab:fit.P}
\begin{tabular}{r|ccc}
\hline\hline\rule[-1.4ex]{0pt}{4.3ex}
$i$ & \multicolumn{3}{c}{$a_i$}\\
  & BSk19 & BSk20 & BSk21  \\
\hline\rule{0pt}{2.7ex}
1  & 3.916 &     4.078 &   4.857 \\
2  & 7.701 &     7.587 &   6.981 \\
3  & 0.00858 &   0.00839 & 0.00706 \\
4  & 0.22114 &   0.21695 & 0.19351 \\
5  & 3.269 &     3.614 &   4.085 \\
6  & 11.964 &    11.942 &  12.065 \\
7  & 13.349 &    13.751 &  10.521 \\
8  & 1.3683 &    1.3373 &  1.5905 \\
9  &  3.254 &    3.606 &   4.104 \\
10 &  $-12.953$ &  $-22.996$ & $-28.726$ \\
11 &  0.9237 &   1.6229 &  2.0845 \\
12 &  6.20 &     4.88 &    4.89 \\
13 &  14.383 &   14.274 &  14.302 \\
14 &  16.693 &   23.560 &  22.881 \\
15 &  $-1.0514$ &  $-1.5564$ & $-1.7690$ \\
16 &  2.486 &    2.095 &   0.989 \\
17 & 15.362 &    15.294 &  15.313 \\
18 & 0.085 &     0.084 &   0.091 \\
19 & 6.23 &      6.36 &    4.68 \\
20 & 11.68 &     11.67 &   11.65 \\
21 & $-0.029$ &    $-0.042$ &  $-0.086$ \\
22 & 20.1 &      14.8 &    10.0 \\
23 & 14.19 &     14.18 &   14.15 \\
\hline\hline
\end{tabular}
\end{table}
For  BSk19, BSk20 and BSk21, following  \cite{potekhin13}, we adopted the analytical form
\begin{eqnarray}
  \zeta &=&
    \frac{a_1+a_2\xi+a_3\xi^3}{1+a_4\,\xi}\,
        \left\{\exp\left[a_5(\xi-a_6)\right]+1\right\}^{-1}
\nonumber\\&&
     + (a_7+a_8\xi)\,
        \left\{\exp\left[a_9(a_6-\xi)\right]+1\right\}^{-1}
\nonumber\\&&
     + (a_{10}+a_{11}\xi)\,
\left\{\exp\left[a_{12}(a_{13}-\xi)\right]+1\right\}^{-1}
\nonumber\\&&
     + (a_{14}+a_{15}\xi)\,
\left\{\exp\left[a_{16}(a_{17}-\xi)\right]+1\right\}^{-1}
\nonumber\\&&
     + \frac{a_{18}}{1+ [a_{19}\,(\xi-a_{20})]^2}
     + \frac{a_{21}}{1+ [a_{22}\,(\xi-a_{23})]^2}
\label{fit.P}
\end{eqnarray}
The values of parameters $a_i$ are given in Table~\ref{tab:fit.P}.

\section*{Acknowledgements}
SC and MDL are supported by INFN ({\it iniziative specifiche} TEONGRAV and QGSKY). This work was partially supported by the Ministry of Education and Science (Russia).

\end{document}